\documentclass[showpacs,aps,prl,amsmath,amssymb,reprint,superscriptaddress,a4paper]{revtex4-1}

\usepackage{graphicx}% Include figure files
\usepackage{dcolumn}% Align table columns on decimal point
\usepackage{bm}% bold math
\usepackage{hyperref}% add hypertext capabilities
\usepackage{color}
\usepackage{natbib}
\begin{document}

%\preprint{}

\title{Fluid depletion in shear bands}

\author{Roman Mani}

\email[Electronic address: ]{manir@student.ethz.ch}
%\thanks{}
\affiliation{Computational Physics, IfB, ETH-H\"onggerberg, Wolfgang-Pauli-Strasse 27, 8093 Z\"urich, Switzerland}
\author{Dirk Kadau}
\affiliation{Computational Physics, IfB, ETH-H\"onggerberg, Wolfgang-Pauli-Strasse 27, 8093 Z\"urich, Switzerland}
\author{Dani Or}
\affiliation{Inst. f. Terrestrische Oekosysteme, ETH-Zentrum, Universit\"atsstrasse 16, 8092 Z\"urich, Switzerland}
\author{Hans J. Herrmann}
\affiliation{Computational Physics, IfB, ETH-H\"onggerberg, Wolfgang-Pauli-Strasse 27, 8093 Z\"urich, Switzerland}


\date{\today}

\begin{abstract}
How does  pore liquid reconfigure within shear bands in wet granular media? Conventional wisdom predicts that liquid is drawn into dilating granular media. We, however, find a depletion of liquid in shear bands despite increased porosity due to dilatancy. This apparent paradox is resolved by a microscale model for liquid transport at low liquid contents induced by rupture and reconfiguration of individual liquid bridges. Measured liquid content profiles show macroscopic depletion bands similar to results of numerical simulations. We derive a modified diffusion description for rupture-induced liquid migration.\end{abstract}

%\keywords{}

\maketitle

Is interstitial pore liquid driven away or drawn into shear bands during deformation of granular matter? 
In fully saturated granular materials it is known that liquid is sucked into dilating shear bands with increase in porosity \cite{Hicher1994,Henkel1956,Viggiani2004,Tillemans1995} since air is precluded from entering the dilated shear band pore volume. In unsaturated granular media, however, at low liquid contents with continuous air phase it is unclear if the liquid content increases or decreases. In this work we show that in the pendular regime where only capillary bridges are present, an opposite liquid migration pattern than observed for saturated media develops. Despite increase in porosity in the shear band, liquid content consistently decreases. This effect is also measured experimentally in a split bottom shear cell where the shear band position is fixed and long shear paths are possible \cite{experiment_split}. Liquid migration is of tremendous importance to the stability properties of soil structures \cite{Book_fredlund,Mitchell_book}. Partially saturated soils may be considered as brittle due to the collapse of menisci in the failure plane \cite{cunningham2003}. Furthermore, granular materials generally lose strength with decreasing liquid content \cite{scheel}. On the other hand, liquid accumulation in soil pores may cause a dramatic decrease in strength leading e.g.\ to landslides or soil collapses \cite{asch1999,Book_fredlund}. It was reported that in partially saturated soils, liquid migrates away from the shear plane due to the formation of microcracks and increasing connectivity near the failure plane \cite{cetin1999}. {The degree of saturation ranged from 45\% to 80\% and the liquid content variations were measured on a macroscopic level by cutting out slices parallel to the shear band.} On the grain level however, first experimental investigations on partially saturated soils using non destructive techniques have been performed only recently \cite{Oka2011} and the authors did not measure the liquid content. Moreover, liquid migration is of great interest also in a variety of other situations in powder technology or pharmaceutical applications where grains are mixed with liquid, e.g.\ in spray coating of tablets \cite{Turton2008}.
\begin{figure}[htbp]
\begin{center}
\includegraphics[width=\columnwidth]{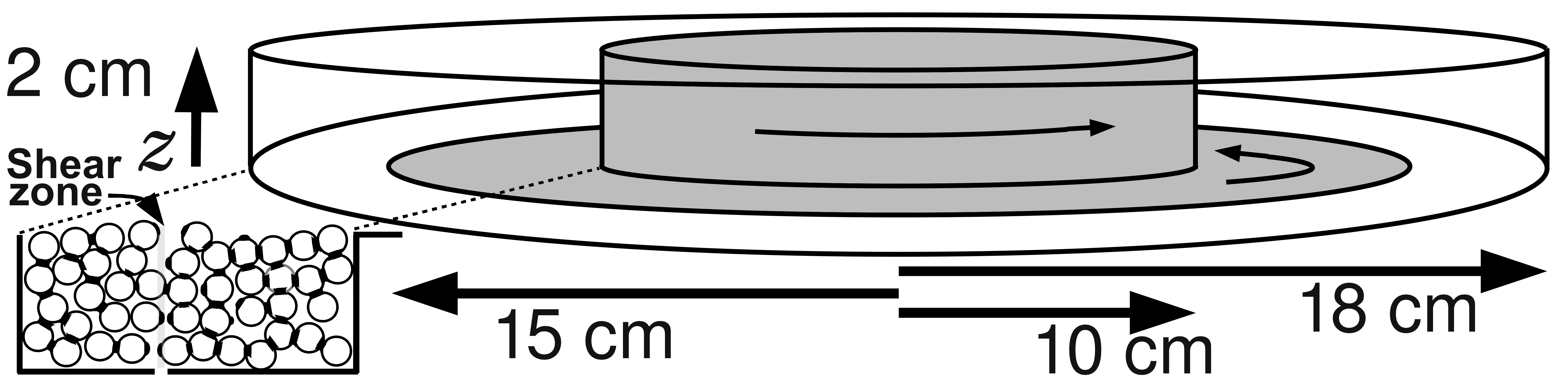} 
\caption{\label{fig:setup}The circular split bottom shear cell consists of a rotating inner part (shaded in gray) as well as a fixed outer part (white) separated by a thin slit. The gray cylinder and ring rotate at the same angular velocity.  }
\end{center}
\end{figure}

\begin{figure}[htbp]
\begin{center}
\includegraphics[width=\columnwidth]{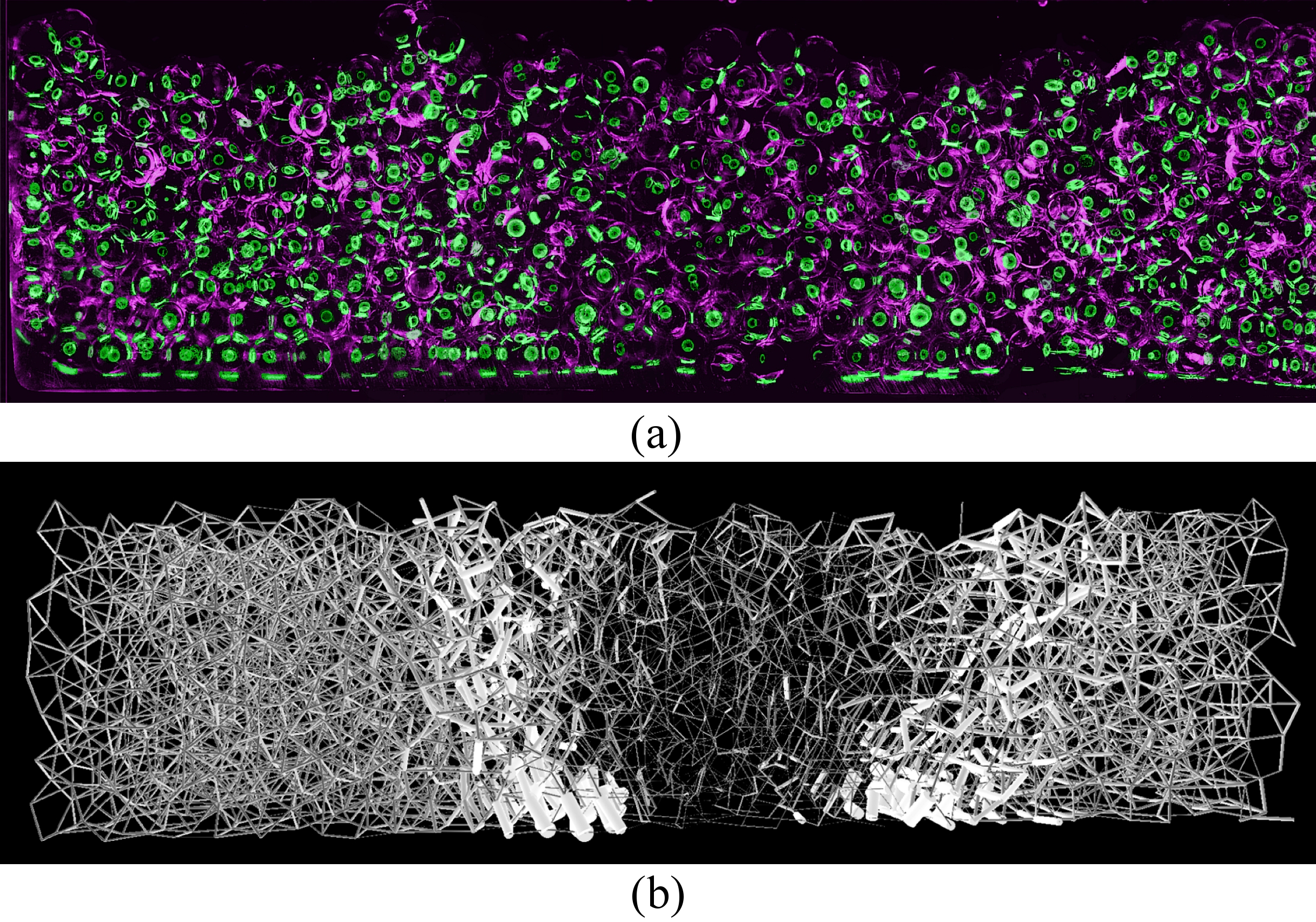} 
\caption{\label{fig:experiment_snap}(Color online) (a) A slice perpendicular to the shear band after a shear displacement of 8~turns. Liquid (artificially highlighted in green) has migrated away from the shear band and is accumulated along the edges of the shear zone. The beads were artificially marked purple. (b) A slice perpendicular to the shear band as obtained in simulations.  We only show the capillary bridges represented by  lines of width and brightness proportional to their volume. The liquid distribution widens with increasing height $z$. The width of both images is 60~mm. }
\end{center}
\end{figure}

\begin{figure}[htbp]
\begin{center}
\includegraphics[width=\columnwidth]{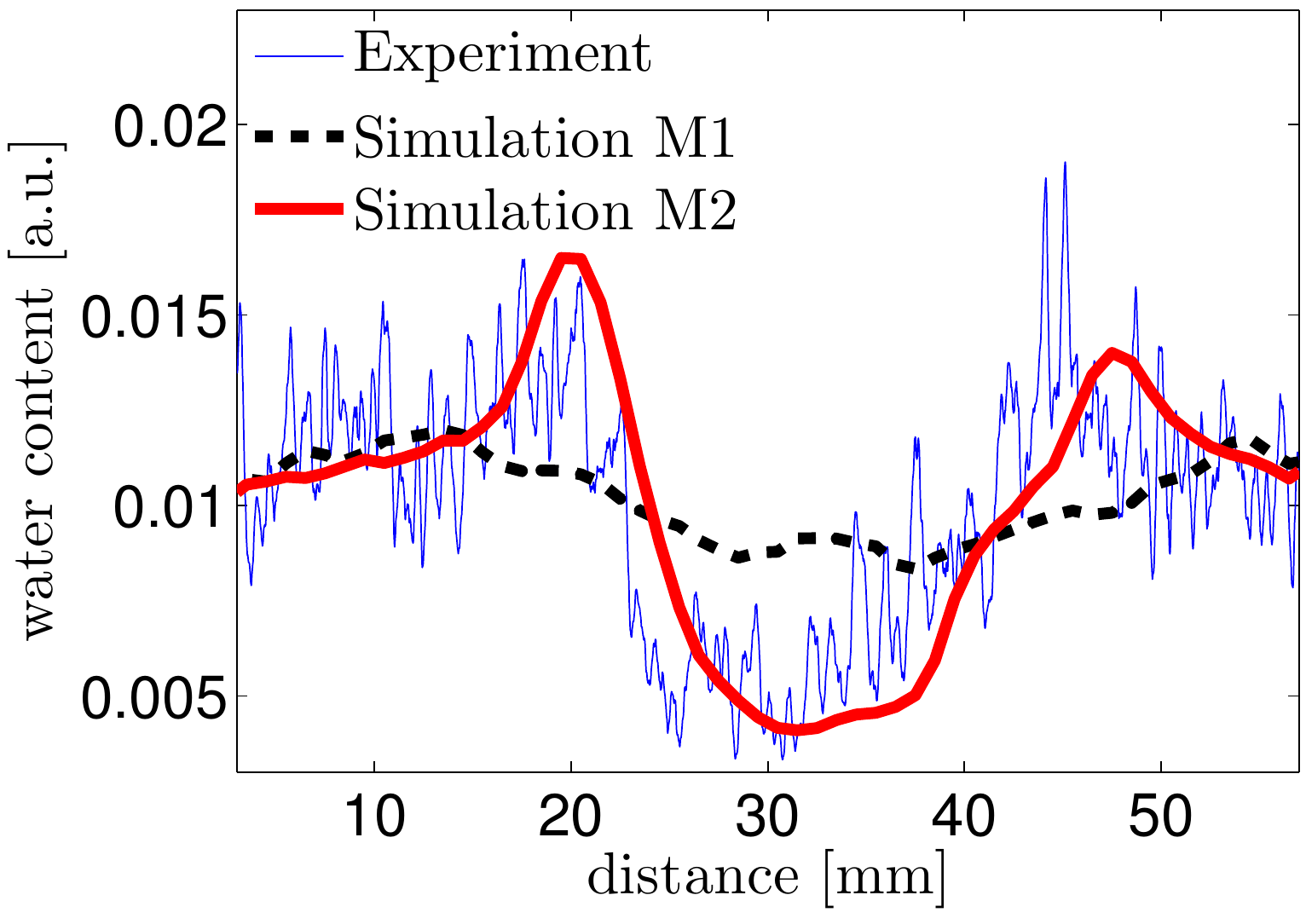}

\caption{\label{fig:experiment_sim_curve}(Color online) The water content from simulation (Simulation M2) and experiment agree very well. A  depletion in humidity is found inside the shear band after 8 turns. Curve Simulation M1 shows the liquid distribution obtained by the suction controlled model.}
\end{center}
\end{figure}

In this Letter, we experimentally measured the liquid content in a shear band and propose a model with capillary bridges capable of reproducing the experimental results.
To investigate liquid transport in regions of strain localization it is desirable to realize a stable shear band at a fixed position. 
The circular split bottom cell \cite{experiment_split} used in our experiments, satisfies this requirement. Our split bottom cell, a sketch is shown in  fig.~\ref{fig:setup}, consists of two circular L shapes where the inner is made of a rotating cylinder at constant velocity (radius 10~cm) and the outer is fixed. The distance from the symmetry axis of the cell to the outer wall is 18~cm and 15~cm to the slit where the L shapes join. We chose a filling height of 1.5~cm and used glass beads (from Sigmund Lindner, SiLi beads type S) with $R=0.85~\mathrm{mm}\pm0.0075~\mathrm{mm}$. We define a coordinate $z$ along the axis of the inner cylinder.  In order to measure the liquid distribution in the bulk of the sample, we used a UV glue (from Norland products, NOA~61) that hardens after irradiation with ultraviolet light. In the liquid state it has a surface tension of 50mN/m which is close to water and a viscosity of 300 mPas. After shearing, the glue is immediately hardened, such that subsequent measurements of the liquid content are possible with virtually no time delay. For visualization we mixed the UV glue with a fluorescent dye and in order to cut the sample into slices after hardening, we filled all remaining pore space with a clear and colorless epoxy resin and checked that the liquid bridges remain in their original state. The rotation speed was fixed to $\omega=2\pi/600s^{-1}$, the overall shear displacement was 8 turns and the liquid content was 1~\% which is the liquid volume divided by the volume of the sample. Fig.~\ref {fig:experiment_snap} (a) shows a slice along a plane perpendicular to the shear band.
Indeed, liquid is driven out of the shear band and is accumulated along the edges. Moreover, the liquid distribution widens with increasing $z$. The thin blue curve in fig.~\ref{fig:experiment_sim_curve} shows the liquid content obtained experimentally as a function of the distance from the outer fixed wall averaged over bridges having a height larger than $z=3.4$~mm. For this particular filling height, the depletion is about 50\%. To determine the liquid content we counted the amount of pixels belonging to a liquid bridge in the images taken by an optical microscope and averaged over 8 slices.

In order to understand the redistribution of liquid in sheared granular flows we develop a model with capillary bridges taking into account the redistribution of liquid after bridge rupture. Using this model we show that liquid is indeed driven out of shear bands.
We use Contact Dynamics\cite{jean_cd,brendel_cd} to model rigid spherical particles with radii uniformly distributed between 0.775~mm and 0.925~mm. The contact forces are calculated based on perfect volume exclusion, Coulomb friction and cohesive forces due to capillary bridges. Simple models for cohesive forces are well established in Contact Dynamics \cite{Kadau03,kadau2009c,taboada2006}. We use empirical formulas for the capillary force which have been suggested by Willet et al.\cite{willet}. 
The grains are characterized by a certain roughness in which liquid may accumulate and form a wetting layer. As soon as liquid layers coating the particles are brought into contact, large negative pressure draws liquid to the contact forming a capillary bridge  at very short time scales of the order of a millisecond for grains of one millimeter diameter  \cite {Herming,PhysRevE.80.031306}. The attractive force exerted by the capillary bridge acts on the two grains until their separation $s$ is larger than a critical rupture distance $s_c\approx V^{1/3}$ \cite{willet}. Usually, slow deformations e.g.\ in geotechnics are studied by means of suction controlled models requiring a constant Laplace pressure in the bridges throughout the sample \cite{Charyere}. Similarly, we implement a suction controlled model by solving the Young-Laplace equation at each contact and require that the total liquid volume is conserved. Using this model we found the liquid content, shown in fig.~\ref{fig:experiment_sim_curve} (curve Simulation M1) to be proportional to the number of contacts only, which is slightly decreased in the shear band. The depletion observed in experiments (thin blue curve) is however signigicantly stronger! Where does this discrepancy come from?

A first hint for implementing a more realistic model is found in studies concerning fluid dynamics in wet granular matter \cite{kohonen,scheel}. It was observed that Laplace pressures are equalized in a finite time of the order of one to five minutes \cite{scheel} for glass beads of half a millimeter diameter. The equalization takes place due to liquid flow driven by Laplace pressure differences either through the vapor phase, which is presently not the case, or through the wetting layers on the beads \cite{scheel}. Primarly by small liquid bridges growing in favor of larger ones. Since the pressure in the bridges is negative, liquid flux between capillary bridges can only occur if liquid pressure in the film is negative, as well. This may happen for thin films due to roughness or due to surface forces. We call this thin film the minimal wetting layer.   
Furthermore, it was shown experimentally, that the associated flow resistance for thin films is essentially independent on the film thickness \cite{seemann2001}. In that sense, the dynamics can be modeled by taking the flux between two bridges as proportional to the pressure difference where the proportionality constant is the inverse resistance. Secondly, there is a dynamic flow process whenever liquid bridges rupture. In that case, the liquid is sucked back onto the grains very fast \cite{Herming} and the film swells considerably. This causes a massive decrease of flow resistance in the film and since the pressure is positive for thick films on bead surfaces, flow takes place on a shorter time scale from the film to the bridges until the film has shrunk back to the minimal wetting layer. 

Based on these considerations we propose the following improved model: Since the volume of the minimal wetting layers on the beads is  small compared to the volume in the bridges we do not consider the volume of the minimal films explicitly. Instead, whenever two beads come into contact, we initialize the liquid bridge with a small volume which is sucked out of the neighboring bridges.  Once a liquid bridge has formed, it becomes part of the contact network in which the volumes are constantly updated according to $\dot{V_i}=\mu\sum_j (P_i-P_j)$ where $P_i$ is the pressure in a contact $i$ and the sum runs over all neighbors $j$ of contact $i$ and $\mu$ is the inverse flow resistance of the film which is a model parameter. Whenever a bridge ruptures, the liquid is equally split among the two particles. Since the film swells (the volume of the film becomes $V_f$) and the flow resistance drops, we assume that the liquid is instantaneously sucked into all neighboring bridges. Each bridge receives an amount of liquid $\Delta V_i=A(P_f-P_i)/L_i$ where $P_f=\Gamma/R$ is the pressure in the film with surface tension $\Gamma$, $L_i$ is the distance from the rupture point to the bridge $i$ and $A$ is a normalization factor such that $\sum_i \Delta V_i=V_f$. 
Here, the total amount of liquid is conserved as opposed to suction controlled models, e.g.\ \cite{Charyere}. Note that the cylinder walls as well as the two bottom rings are considered to be completely hydrophobic. 

Using this model we simulate shear flow in the circular split bottom shear cell with the same dimensions as in the experiment. However, since the system is rotationally invariant with respect to the $z$-axis we only consider a sector of the cell with arc length $\vartheta=0.0873$. We use periodic boundary conditions in the $\phi$-direction, where in cylindrical coordinates, $\phi$ is the angular coordinate. 
Initially, all bridges have the same amount of liquid and we record the evolution of the liquid in the sample during shear. 
Fig. \ref{fig:experiment_snap} (b) shows a snapshot of the capillary bridge network after the same shear displacement as in experiments. The liquid distribution widens along $z$ as in the experimental result.
Curve Simulation M2 in Fig.~\ref{fig:experiment_sim_curve} shows the liquid distribution obtained using this model with $\mu=0$ averaged over 5 independent runs in good agreement with the experimental data. Why could the flow resistance $\mu^{-1}$ be set so small in the simulation? As previously mentioned, equilibration of liquid bridges for water takes about 5 min. However, the viscosity of the UV glue is 300 times larger and the surface tension is 1.5 times smaller than that of water such that the expected equilibration time is about 7 hours which is very large compared to the shearing velocity. This means that the liquid structures do not have enough time to equilibrate, the flow takes place mainly via bridge rupture events leading to smaller bridge volumes inside the shear band. Decreasing the shearing velocity results in a smaller depletion and in the limit $\omega\to0$ we eventually recover curve Simulation M1 in fig~\ref{fig:experiment_sim_curve}.

We now explain this behavior in terms of a theoretical model. For simplicity we first consider a one dimensional shear rate profile which e.g.\ occurs in plane shear flows between two parallel walls separated by some distance along the $z$-direction. Here, the local shear rate depends only on the $z$-coordinate. We divide the plane shear geometry into slices of thickness $h$ along the $z$-axis. Since $\mu$ is small in our experimental system the main transport happens via rupture of individual capillary bridges. If pressure differences are not too large, there is on average an isotropic transport of liquid away from the rupture point which is proportional to the volume of the ruptured bridge. Thus, the amount of liquid leaving a slice of our model plane shear geometry during a time interval $\Delta t$ is proportional to the bridge rupture rate and to the average bridge volume in the slice. The same holds for the neighboring slices $i-1$ and $i+1$ such that the total change of liquid in slice $i$ can be written as
\begin{equation}\label{discrete}
Q^i(t+\Delta t)-Q^i(t)=\frac{\chi \Delta t }{2}(B^{i-1}Q_b^{i-1}+B^{i+1}Q_b^{i+1}-2B^iQ_b^i)
\end{equation}
where $Q_i$ is the total amount of liquid, $B_i$ the bridge rupture rate, $Q_b^i$ the average bridge volume in slice $i$ and $\chi$ is a geometrical proportionality factor which measures the average amount of liquid leaving a slice after each rupture event. On length scales larger than $h$ and $\Delta t\to 0$ we arrive to the following continuum equation
\begin{equation}
\label {continuum}
\dot{Q_b}=C\frac{\partial^2}{\partial z^2}(\dot{\gamma}Q_b)
\end{equation}
where $C$ is a constant and $Q_b$ is the average bridge volume. Since the bridge rupture rate is proportional to the shear rate $\dot\gamma$ and to the number of contacts, we replaced $B$ by $\dot{\gamma}$ and assumed that the number of contacts varies much less with $z$ than the shear rate.  The particles themselves behave diffusively in plane shear flow \cite{campbell_diffusion}. Therefore, the liquid bridges also diffuse in space which gives rise to an additional diffusive contribution $\dot Q_b \sim (\dot\gamma Q_b)^{\prime\prime}$ to eq. (\ref{continuum}) which could be lumped into the constant $C$.

A nice feature of eq.~\ref{continuum} is that for constant $\dot\gamma$, the liquid spreads diffusively as observed in a variety of industrial processes, e.g.\ in tablet spray coating \cite{Turton2008}. However, eq. (\ref{continuum}) is not an ordinary diffusion equation as $\dot\gamma$ is not a global constant. Indeed, the equation predicts that even homogeneous liquid distributions will change if the second derivative of the shear rate profile $\dot \gamma^{\prime\prime}$ with respect to $z$ is large. We now address the liquid depletion pattern observed (see fig.~\ref{fig:experiment_snap}): The shear rate profile in split bottom shear cell geometries is Gaussian \cite{PhysRevE.76.051301,experiment_split} and its width $W(z)$ increases as a function of $z$. Therefore, at fixed $z$, $\dot\gamma ^{\prime\prime}$ is smallest and negative in the center of the shear band which causes a liquid content drop due to large fluxes away from the center induced by rupture of individual bridges. However it is largest and positive at the edges which explains the accumulation of liquid along the edges of the shear band.

\begin{figure}[htbp]
\begin{center}

\includegraphics[width=\columnwidth]{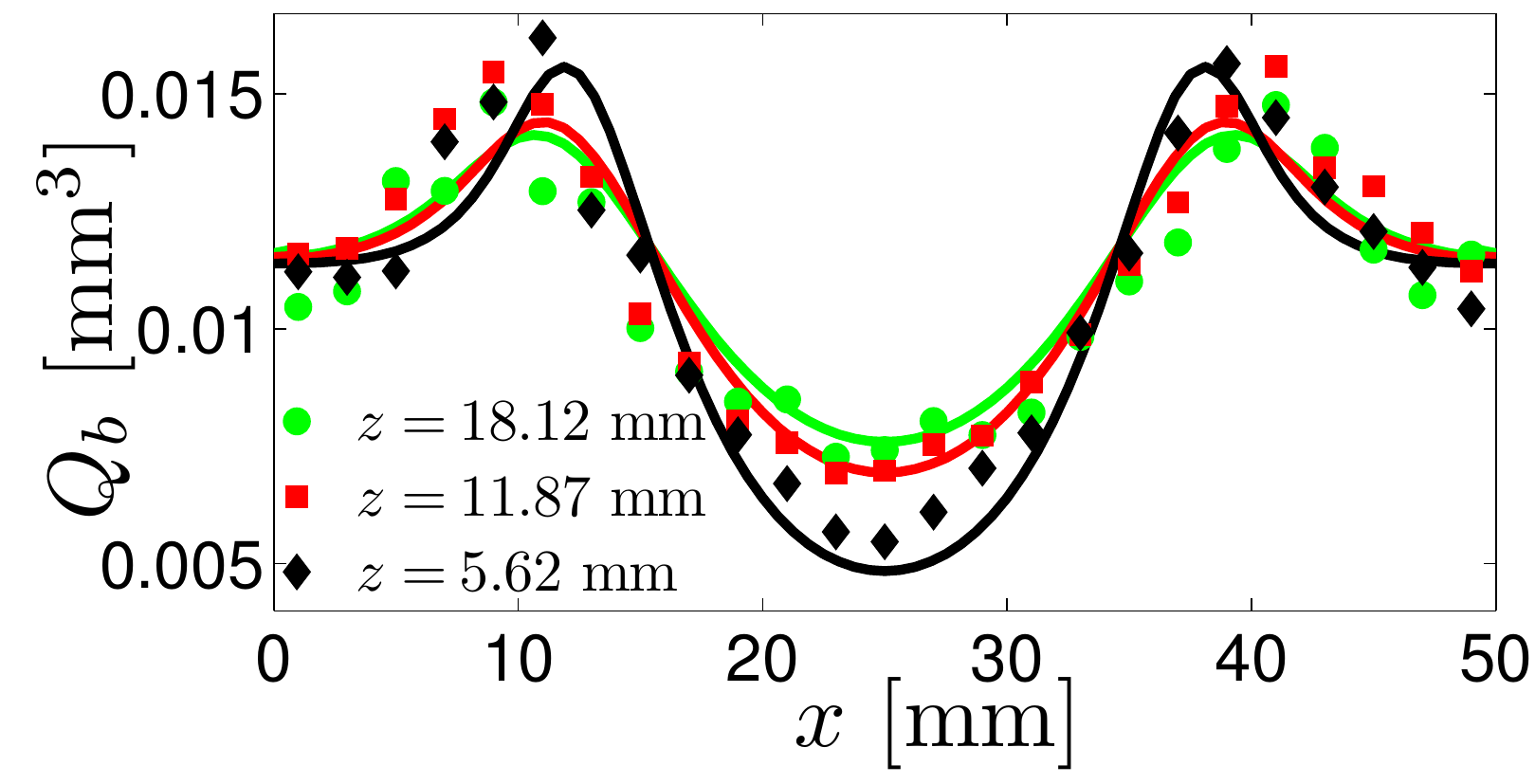}
\caption{\label{fig:split_sim}(Color online) The average bridge volume $Q_b$ in the linear split bottom shear cell is plotted for different $z$ as a function of $x$ after a shear displacement of 8 m. The data from simulations (symbols) agrees well with the numerical solution (lines) of eq. (\ref{continuum}).}
\end{center}
\end{figure}

In order to validate the proposed model we simulate the simpler linear split bottom shear cell which consists of two straight L-shapes sliding past each other and we compare to the numerical solution of eq.~\ref{continuum}. We now use cartesian coordinates where the $x$-direction is perpendicular and the $y$-direction is parallel to the slit and the $z$-direction is perpendicular to the bottom plates. To ensure a homogenous sample, we add a frictionless top plate on which a force is exerted to confine the grains in the container and we switch off gravity as well as cohesion. The dimensions of the system are now $L_x=50$~mm, $L_y=18$~mm and $L_z\sim19$~mm and bead radii uniformly distributed between 0.735~mm and 0.925~mm.

Our system is effectively two dimensional, since it depends on both, $x$ and $z$. Thus, we generalize Eq. (\ref{continuum}) to $\dot Q=C\triangle (|\dot\gamma| Q)$ where $|\dot{\gamma}|$ is the second invariant of the shear rate tensor \cite{jop}. This equation is solved over the domain $z>0.625$~mm by using Neumann boundary conditions, the same initial condition and shear rate as in the simulation and $C=0.0675$~$\mathrm{mm}^2$ . Fig.~\ref{fig:split_sim} shows the simulation results for $Q_b$ as a function of $x$ and different $z$ (symbols) which are in good agreement with the numerical solution of the previous equation (lines). Only for small $z$, the results agree less well. We attribute this to the fact that the shear rate near the bottom rings varies on small length scales which is not resolved in the averages of the simulations. Note that although the system is in mechanical steady state, the liquid distribution still changes on longer time scales until $|\dot\gamma|Q$ is spatially constant. According to eq.~\ref{continuum} the driving of liquid at the beginning of shearing is inversely proportional to the width squared of the shear rate profile! We noticed that for our system, rather large shear displacements lead to considerable liquid migration. However, shear bands occurring in nature, e.g. in soils can be very narrow such that the strains needed to dry the shear band will be much smaller than presented in this work.

In conclusion, we found that liquid amount decreases within unsaturated shear bands at low liquid contents.  A  simple model for liquid redistribution explains this discovery. The driving mechanism is relative motion of grains leading to bridge rupture and
associated liquid transport. We  derived a modified diffusion equation describing liquid migration within the system and that was verified experimentally. Our work shows that knowing how liquid is redistributed due to shear is crucial for prolonged simulations and experiments in the field of wet granular matter e.g.\ in rheological measurements where long lived shear banding is omnipresent.  Especially at high shear rates, where liquid depletion and loss of capillary bridges may induce noticeable shear softening and thixotropic effects.

\begin{acknowledgments}
We would like to thank Martin Brinkmann for helpful discussions. We greatly acknowledge  the provision of a lookup table allowing the determination of the Laplace pressure by Ciro Semprebon as well as technical support by Daniel Breitenstein and Gabriele Peschke. We thank the Deutsche Forschungsgemeinschaft (DFG) for financial support through grant No. HE 2732/11-1.
\end{acknowledgments}

\bibliography{references}

\end{document}